\documentclass[AMA,STIX2COL,Linenumbersoff]{MRM}
\articletype{Research Article}%

\received{dd mm yyyy}
\revised{dd mm yyyy}
\accepted{dd mm yyyy}
\begin{document}

\title{Utilizing 3D Fast Spin Echo Anatomical Imaging to Reduce the Number of Contrast Preparations in \texorpdfstring{$T_{1\rho}$}{T1rho} Quantification of Knee Cartilage Using Learning-Based Methods}

\author[1]{Junru Zhong}{\orcid{0000-0002-3897-9280}}
\author[1]{Chaoxing Huang}{\orcid{0000-0003-4941-5840}}
\author[1]{Ziqiang Yu}{\orcid{0009-0000-2466-4942}}
\author[2]{Fan Xiao}{\orcid{0009-0002-3766-6917}}
\author[1]{Siyue Li}{\orcid{0000-0001-8791-5841}}
\author[3]{Tim-Yun Michael Ong}{\orcid{0000-0002-4460-9286}}
\author[4]{Ki-Wai Kevin Ho}{\orcid{0000-0001-8647-8475}}
\author[5]{Queenie Chan}{\orcid{0000-0003-1270-995X}}
\author[1]{James F. Griffith}{\orcid{0000-0002-4460-9286}}
\author[1]{Weitian Chen}{\orcid{0000-0001-7242-9285}}

\authormark{ZHONG \textsc{et al}}

\address[1]{\orgdiv{CU Lab of AI in Radiology (CLAIR), Department of Imaging and Interventional Radiology}, \orgname{The Chinese University of Hong Kong}, \orgaddress{\state{Hong Kong SAR}, \country{China}}}

\address[2]{\orgdiv{Department of Radiology}, \orgname{Shanghai Sixth People’s Hospital Affiliated to Shanghai Jiao Tong University School of Medicine}, \orgaddress{\state{Shanghai}, \country{China}}}

\address[3]{\orgdiv{Department of Orthopaedics \& Traumatology}, \orgname{The Chinese University of Hong Kong}, \orgaddress{\state{Hong Kong SAR}, \country{China}}}

\address[4]{\orgdiv{Department of Orthopaedics \& Traumatology}, \orgname{The Chinese University of Hong Kong Medical Centre}, \orgaddress{\state{Hong Kong SAR}, \country{China}}}

\address[5]{\orgname{Philips Healthcare}, \orgaddress{\state{Hong Kong SAR}, \country{China}}}

\corres{Corresponding to Weitian Chen, Room 15, LG/F, Cancer Centre, Prince of Wales Hospital, Sha Tin, NT, Hong Kong SAR, China \email{wtchen@cuhk.edu.hk}}

\finfo{This work was supported by \fundingAgency{Innovation and Technology Commision of HKSAR Government} grant \fundingNumber{MRP/001/18X}}

\abstract[Summary]{

\section{Purpose}
To propose and evaluate an accelerated $T_{1\rho}$ quantification method that combines $T_{1\rho}$-weighted fast spin echo (FSE) images and proton density (PD)-weighted anatomical FSE images, leveraging deep learning models for $T_{1\rho}$ mapping. The goal is to reduce scan time and facilitate integration into routine clinical workflows for osteoarthritis (OA) assessment.

\section{Methods}
This retrospective study utilized MRI data from 40 participants (30 OA patients and 10 healthy volunteers). A volume of PD-weighted anatomical FSE images and a volume of $T_{1\rho}$-weighted images acquired at a non-zero spin-lock time were used as input to train deep learning models, including a 2D U-Net and a multi-layer perceptron (MLP). $T_{1\rho}$ maps generated by these models were compared with ground truth maps derived from a traditional non-linear least squares (NLLS) fitting method using four $T_{1\rho}$-weighted images. Evaluation metrics included mean absolute error (MAE), mean absolute percentage error (MAPE), regional error (RE), and regional percentage error (RPE).

\section{Results}
Deep learning models achieved RPEs below 5\% across all evaluated scenarios, outperforming NLLS methods, especially in low signal-to-noise conditions. The best results were obtained using the 2D U-Net, which effectively leveraged spatial information for accurate $T_{1\rho}$ fitting. The proposed method demonstrated compatibility with shorter TSLs, alleviating RF hardware and specific absorption rate (SAR) limitations.

\section{Conclusion}
The proposed approach enables efficient $T_{1\rho}$ mapping using PD-weighted anatomical images, reducing scan time while maintaining clinical standards. This method has the potential to facilitate the integration of quantitative MRI techniques into routine clinical practice, benefiting OA diagnosis and monitoring.
}


\wordcount{4373}

\jnlcitation{\cname{%
\author{J. Zhong}, 
\author{C. Huang}, 
\author{Z. Yu}, 
\author{F. Xiao},
\author{S. Li},
\author{TYM. Ong},
\author{KWK. Ho},
\author{Q. Chan},
\author{JF. Griffith},
and 
\author{W. Chen}} (\cyear{2025}), 
\ctitle{Utilizing 3D Fast Spin Echo Anatomical Imaging to Reduce the Number of Contrast Preparations in \texorpdfstring{$T_{1\rho}$}{T1rho} Quantification of Knee Cartilage Using Learning-Based Methods}, \cjournal{Magn. Reson. Med.}, \cvol{xxxxxx}.}

\maketitle

\section{Introduction}\label{intro}

Spin-lattice relaxation time in the rotating frame ($T_{1\rho}$) imaging is an advanced MRI technique for evaluating cartilage composition \cite{roemerImagingOsteoarthritis2022, chalianQIBAProfileMRIbased2021}. Studies have shown that proteoglycan loss in cartilage is associated with increased $T_{1\rho}$ values \cite{mlynarikRoleRelaxationTimes1999, akellaProteoglycaninducedChangesT1rrelaxation2001a}. This method can be performed on standard 1.5T and 3T MRI scanners without requiring specialized hardware or contrast agents, making it a promising tool for the early detection of osteoarthritis (OA) and the monitoring of cartilage therapies in clinical practice.
Extended scan time is a major challenge in $T_{1\rho}$ quantification. For example, the knee $T_{1\rho}$ imaging protocol recommended by the Radiological Society of North America (RSNA) requires approximately 6 to 12 minutes \cite{chalianQIBAProfileMRIbased2021}. This is because multiple $T_{1\rho}$-weighted images must be acquired at the same location to fit a $T_{1\rho}$ relaxation model and calculate $T_{1\rho}$ maps. To address this, many studies have investigated methods to accelerate $T_{1\rho}$ imaging, such as k-space undersampling or reducing the number of $T_{1\rho}$-weighted images \cite{zibettiAcceleratedMonoBiexponential2020, liSuperMAPDeepUltrafast2023, zhangCramerRaoBoundinformed2022, maoBiasreducedNeuralNetworks2024,huangBreathingFreelySelfsupervised2022, huangUncertaintyawareSelfsupervisedNeural2022}.

The two-parameter mono-exponential relaxation model is commonly used for $T_{1\rho}$ quantification. This model requires a minimum of two $T_{1\rho}$-weighted images, though four images are often recommended to ensure robust quantification \cite{chalianQIBAProfileMRIbased2021}. Reducing the number of $T_{1\rho}$-weighted images increases sensitivity to noise, necessitating a high signal-to-noise ratio (SNR) in the acquired data. Recent advances in deep learning have shown that reliable $T_{1\rho}$ mapping can be achieved using only two $T_{1\rho}$-weighted images \cite{huangBreathingFreelySelfsupervised2022, huangUncertaintyawareSelfsupervisedNeural2022}. These methods leverage large training datasets and underlying signal models to enable robust $T_{1\rho}$ fitting even in low-SNR conditions. In this context, deep learning neural networks act as approximations of the $T_{1\rho}$ signal equation \cite{hornikMultilayerFeedforwardNetworks1989}, offering strong noise tolerance. Inspired by the representational capabilities of deep learning, we hypothesized that $T_{1\rho}$ imaging could be further accelerated through an optimized acquisition strategy. For instance, $T_{1\rho}$ fitting could be achieved using just two images: one $T_{1\rho}$-weighted image and one image acquired with a standard clinical pulse sequence.

In this study, we examined the feasibility of generating $T_{1\rho}$ maps using a proton-density (PD)-weighted anatomical image and a single $T_{1\rho}$-weighted image, both acquired with fast/turbo spin echo (FSE/TSE) sequences. The $T_{1\rho}$-weighted image was obtained using a $T_{1\rho}$-prepared FSE sequence \cite{jordanVariabilityCubeQuantT1rho2014, chen3DQuantitativeImaging2011}, while the PD-weighted FSE acquisition served as a surrogate for image contrast corresponding to a time-of-spin-lock (TSL) of zero in conventional quantitative $T_{1\rho}$ imaging. In our experiments, we assessed the performance of the proposed method under various acquisition parameters and compared it to the commonly used non-linear least-squares (NLLS) fitting approach.

\section{Methods}\label{sec:methods}

\subsection{MRI Pulse Sequence}
In accordance with previous studies \cite{jordanVariabilityCubeQuantT1rho2014, chenErrorsQuantitativeT1rho2015, chen3DQuantitativeImaging2011, chenBreathholdBlackBlood2016}, we employed a magnetization-prepared 3D FSE acquisition in this study. The pulse sequence initiates with a magnetization reset, followed by a $T_1$ recovery period, a spin-lock preparation module, and a 3D FSE readout. SPectral Attenuated Inversion Recovery (SPAIR) was interoperated during the $T_1$ recovery period for fat suppression. The spin-lock module consists of a 90-degree tip-down RF pulse to tip magnetization into the transverse plane, succeeded by spin-lock RF pulse clusters with a duration TSL and amplitude at the frequency of spin-lock (FSL), followed by a 90-degree tip-up RF pulse to flip magnetization to the longitudinal direction. The $T_{1\rho}$-preparation is compensated for $B_1$ and $B_0$  \cite{witscheyArtifactsT1rweightedImaging2007}. Following the $T_{1\rho}$ preparation, imaging data is acquired using a 3D FSE readout with a short echo time (TE) and centric view ordering. To minimize potential artifacts caused by rapid signal variations at the start of the FSE readout, a few echoes at the beginning of the echo trains are omitted. With the aforementioned pulse sequence design, the magnetization due to $T_{1\rho}$ relaxation follows the signal equation below:

\begin{equation}\label{eq:t1rho}
    I_k = I_0 e^{-\frac{1}{T_{1\rho}} TSL_k}
\end{equation}

where $I_0$ and $I_k$ are the magnitude of the $T_{1\rho}$-weighted images acquired with TSL=0ms and TSL=$TSL_k$, respectively. 

It is noteworthy that with such a pulse sequence design, the $T_{1\rho}$-weighted images acquired at TSL=0ms have contrast comparable to conventional proton density (PD)-weighted images. Consequently, we hypothesize that a 3D PD-weighted anatomical FSE image, combined with a single $T_{1\rho}$-weighted image acquired at a non-zero TSL, can be utilized to achieve simultaneous $T_{1\rho}$ quantification and anatomical imaging.

\subsection{Data Acquisition}

\begin{table*}[t]%
\caption{MRI Acquisition Parameters. \label{tab:mri-params}}
\begin{tabular*}{\textwidth}{@{\extracolsep\fill}lcc@{\extracolsep\fill}}
\toprule
\textbf{Parameter} & \textbf{$T_{1\rho}$}  & \textbf{PD-weighted FSE} \\
\midrule
Plane & Sagittal & Sagittal \\
Fat suppression & SPAIR\tnote{$^{1}$} & SPAIR \\
No. of slices & 44 & 292 \\
Field of view ($mm^3$) & 160 $\times$ 160 $\times$ 132 & 130 $\times$ 150 $\times$ 161 \\
TE\tnote{$^{2}$}/TR\tnote{$^{3}$} (ms) & 31/2000 & 30/1200 \\
Resolution ($mm^3$) & 0.8 $\times$ 1 $\times$ 3 & 0.55 $\times$ 0.545 $\times$ 0.55 \\
Spin-lock frequency (Hz) & 300 & N/A\tnote{$^{4}$} \\
Spin-lock time (ms, in acquisition order) & 0/50/30/10 & N/A \\
Scan time (min: sec) & 4:02 & 7:20 \\
\bottomrule
\end{tabular*}
\begin{tablenotes}%%[341pt]
\item[$^{1}$] SPAIR = SPectral Attenuated Inversion Recovery
\item[$^{2}$] TE = echo time
\item[$^{3}$] TR = repetition time
\item[$^{4}$] N/A = Not Applicable
\end{tablenotes}
\end{table*}

We retrospectively conducted our \textit{in vivo} experiments on a previously reported dataset \cite{zhongSystematicPostProcessingApproach2024}. Our study received approval from the institutional review board, and all participants provided informed consent. The dataset comprised 40 participants (30 OA patients and 10 healthy volunteers), with a mean age of 56.4$\pm$19.9 years (mean$\pm$standard deviation) and a mean body mass index (BMI) of 24.7$\pm$4.2 kg/m$^2$ (mean$\pm$standard deviation). 14 (35.00\%) of the participants were male.

For every participant, four 3D volumes of $T_{1\rho}$-weighted images were acquired with an FSL of 300 Hz and TSL values of 0, 10, 30, 50ms, respectively, using a 3T clinical MRI scanner (Achieva, Philips Healthcare, Best, Netherlands). Additionally, we collected a 3D volume of PD-weighted, fat-suppressed anatomical FSE images. A single volume of $T_{1\rho}$-weighted images were retrospectively selected, along with the volume of PD-weighted anatomical images, to estimate the $T_{1\rho}$ map using the proposed learning-based method. The $T_{1\rho}$ maps estimated from all four $T_{1\rho}$-weighted images with TSL 0, 10, 30, 50ms using an NLLS fitting method were used as the ground truth. Detailed MRI acquisition parameters are provided in Table \ref{tab:mri-params}. 

\subsection{\texorpdfstring{$T_{1\rho}$}{T1rho} Fitting}

In this section, we introduce the deep learning-based neural networks that fit $T_{1\rho}$ from the acquired data and accompanying preprocessing.

\subsubsection{Preprocessing}\label{sec:preproc}

We segment the femoral, tibial, and patellar cartilage into one unified cartilage region of interest (ROI) on the $T_{1\rho}$-weighted images. Concurrently, we registered the PD-weighted FSE images with the $T_{1\rho}$-weighted images. It is important to note that each pair of PD-weighted and $T_{1\rho}$-weighted images must be collected from the same subject.
The registration process was executed in three stages: rigid, affine, and symmetric deformable with a validated method \cite{avantsSymmetricDiffeomorphicImage2008}. Registration is crucial in the experimental framework due to the significant differences between the two sequences used for acquisition. Following registration, all images — including the PD-weighted and $T_{1\rho}$-weighted images — underwent Gaussian smoothing with a radius of three to minimize noise. In certain experiments, we further processed the data using ROI masks. Specifically, only the regions within the ROI were retained, while the voxels outside the ROI were set to 0. These experiments are detailed in Section \ref{sec:exp-mask-roi}.

\subsubsection{Deep Learning Fitting Models}

We developed two neural network architectures specifically designed for this task: a 2D U-Net with an output range limiter and a multi-layer perceptron (MLP) model incorporating skip connections. To compare the efficacy of these two models, we conducted experiments as described in Section \ref{sec:exp-choice-nn}.

Given our limited sample size, all trainings and validations were executed using five-fold cross-validation to ensure robustness. The same five-fold split was used across all experiments to ensure a unified evaluation.

\paragraph{2D U-Net Architecture}

\begin{figure*}
\centerline{\includegraphics[width=1\linewidth]{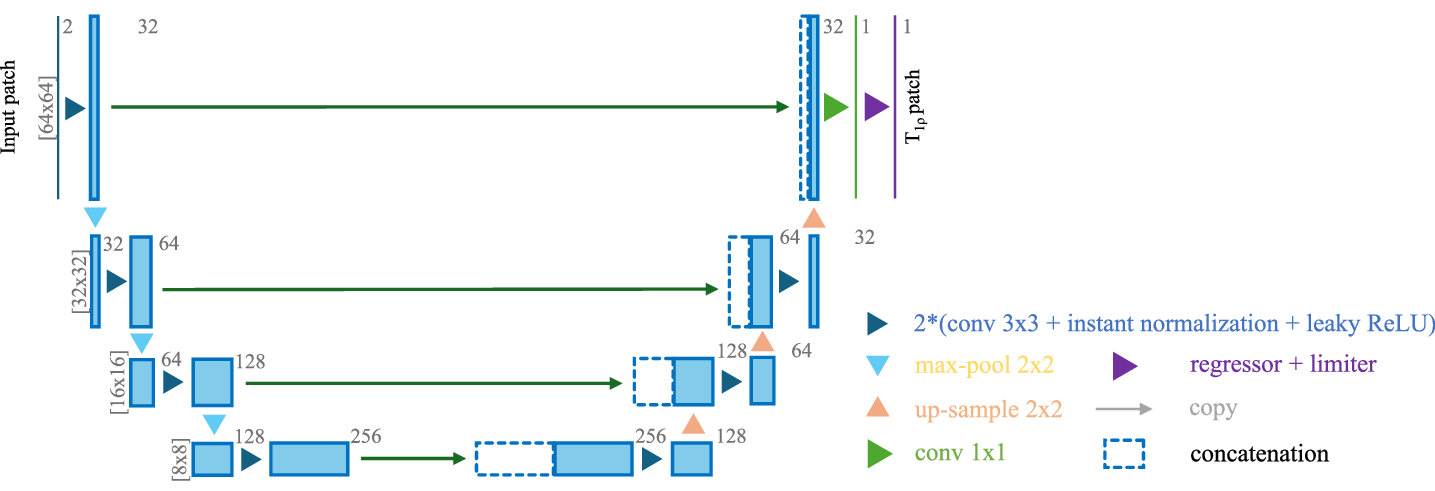}}
\caption{Architecture of the 2D U-Net model. Note, conv = convolution layer.\label{fig:unet}}
\end{figure*}

\begin{figure}
\centerline{\includegraphics[width=1\linewidth]{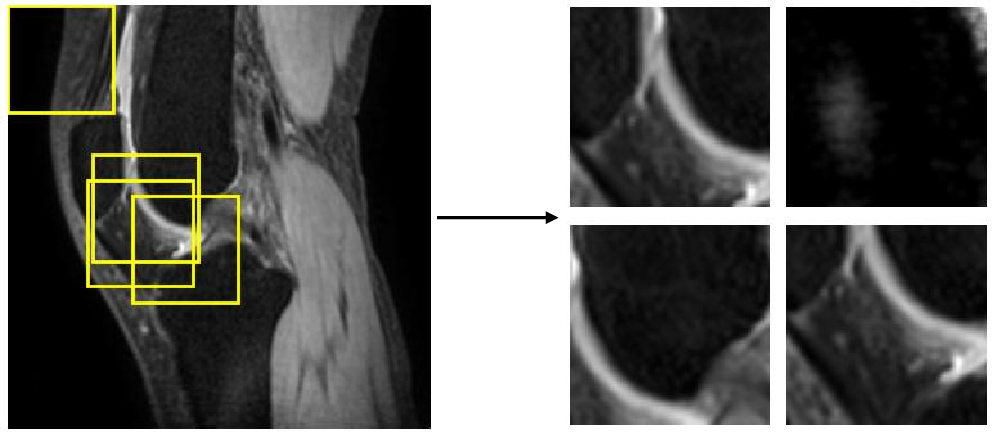}}
\caption{An example slice illustrating the random patching technique employed in 2D U-Net. The yellow boxes indicate the patch positions. This slice was taken from $T_{1\rho}$-weighted image (TSL=0ms) of a healthy volunteer (case V014, 22 years female, BMI = 20.76 kg/m$^2$). Note, ROI = region of interest, TSL = spin lock time, BMI = body mass index. \label{fig:patching}}
\end{figure}

Figure \ref{fig:unet} illustrates the architecture of the 2D U-Net model structure. We adopted a standard U-Net structure \cite{ronnebergerUNetConvolutionalNetworks2015} with modifications to accommodate our fitting task by incorporating a regressor and a limiter. The regressor generated a continuous $T_{1\rho}$ prediction, while the limiter constrains the gradient of mispredicted $T_{1\rho}$ value.

The limiter first applied ReLU activation \cite{agarapDeepLearningUsing2019} to eliminate the negative inputs. Subsequently, we biased the ReLU output with the minimum value and clamped the final output. We formulate this limiter in Equation \ref{eq:limiter}.

\begin{equation}\label{eq:limiter}
    \hat{y} = \{y_{min}, ReLU(x) + y_{min}, y_{max}\}
\end{equation}

where $\hat{y}$ represents the final prediction, $x$ denotes the output of the regressor, and $y_{min}$ and $y_{max}$ are hyperparameters to regulate the range of the prediction. These values were set to 10 and 100 based on our prior experience. By incorporating domain knowledge through the limiter, we facilitate more rapid convergence.

The 2D U-Net model was trained using the ground truth $T_{1\rho}$ maps and an L1 (mean absolute error, MAE) loss function. Training occurred over 1000 epochs using the Adam optimizer \cite{kingmaAdamMethodStochastic2017} with an initial learning rate of 0.001, which was exponentially decayed by a factor of 0.9 as the training progressed.

The 2D U-Net model was fed with patched data with dimensions set to 64 $\times$ 64 pixels. Illustrated in Figure \ref{fig:patching}, patching enhanced the visibility of the cartilage ROI, which is relatively small compared to the entire slice. During training, the patches were randomly cropped from 2D slices, with a higher probability of selecting regions around the cartilage ROI. Augmentations were applied to the patches, including random flip, rotation, translation, and Gaussian noise addition, to enhance data diversity and prevent overfitting. When testing, we employed a sliding window strategy (window size 64 $\times$ 64 pixels) to feed every pixel from the slices containing a cartilage ROI into the 2D U-Net model. We further compiled the output slices of $T_{1\rho}$ predictions to form 3D volumes based on their spatial positions for unified volume-based statistical analysis.

\paragraph{1D MLP Architecture}

\begin{figure*}
\centerline{\includegraphics[width=0.75\linewidth]{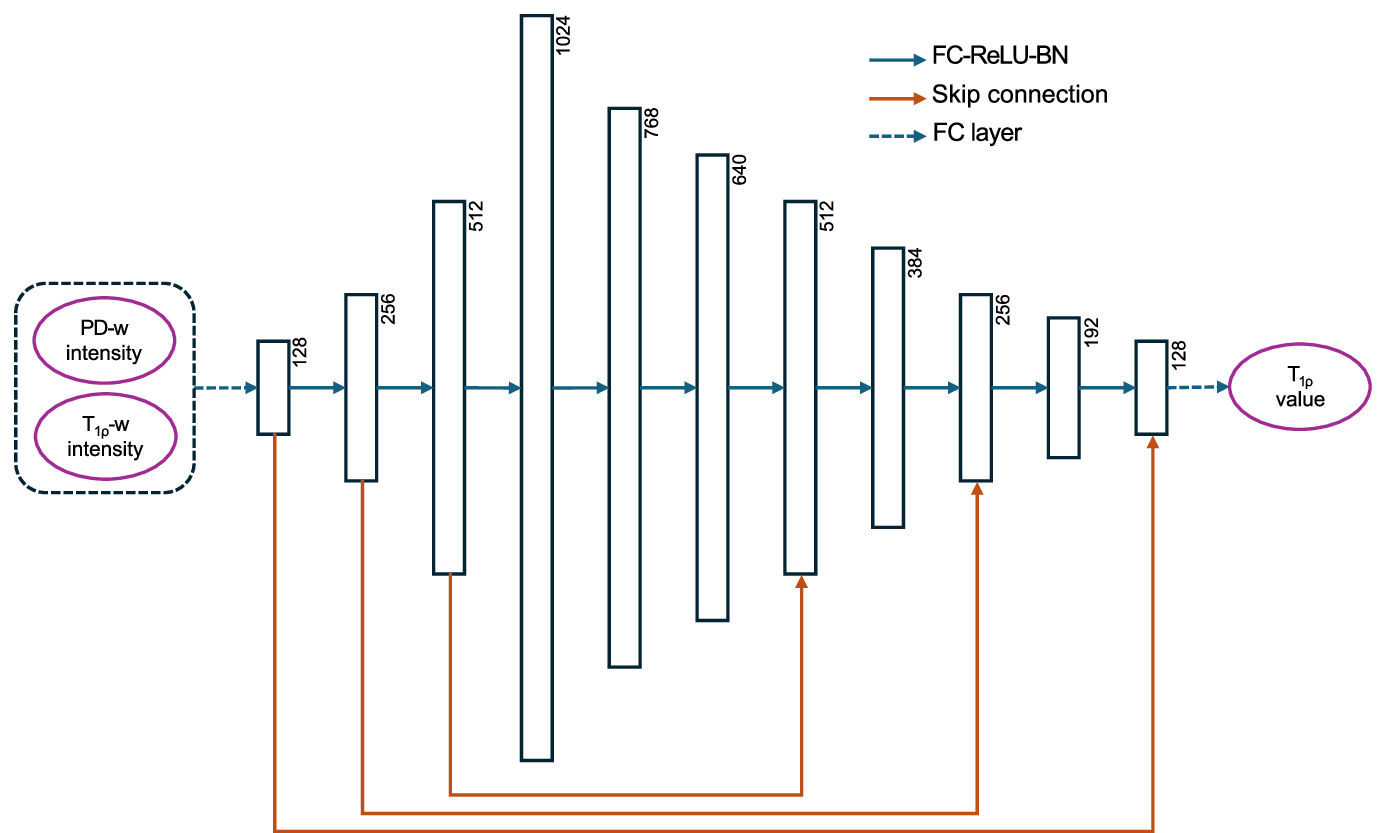}}
\caption{Architecture of the 1D MLP model. The PD-weighted and $T_{1\rho}$-weighted images were concatenated and input to the model in two channels. Legend \textit{FC-ReLU-BN} means three layers in this order. Note, MLP = multi layer perceptron, PD-w = proton density-weighted, $T_{1\rho}$-w = $T_{1\rho}$-weighted, FC = fully connected, BN = batch normalization.\label{fig:mlp}}
\end{figure*}

As illustrated in Figure \ref{fig:mlp}, the MLP architecture was adapted from Zhang et al. \cite{zhangCramerRaoBoundinformed2022}. This model accepts voxel intensities as the input and outputs the corresponding $T_{1\rho}$ values. We extracted voxel intensities from the cartilage ROI in the preprocessed images and ground truth $T_{1\rho}$ to form 1D vectors. These 1D vectors were subsequently fed into the MLP model for training. During evaluation, the 1D $T_{1\rho}$ prediction vectors from the MLP model were reconstructed back into 3D volumes based on the voxel positions. This extraction process ensured a unified, volume-based statistical analysis across all methods and experiments.

The MLP model was trained for 1000 epochs with a batch size of 512. It was optimized using L1 loss and the RMSProp optimizer \cite{gravesGeneratingSequencesRecurrent2014} (initial learning rate = 0.001, weight decay = 0.0003), with exponential learning rate decay at a rate of 0.9.

\subsection{Experiment Design}

We designed our experiments to address three key questions below.

\begin{itemize}
    \item How do the selection between PD-weighted and baseline $T_{1\rho}$-weighted (TSL=0) images, and the choice of TSL in $T_{1\rho}$ imaging, affect the precision of $T_{1\rho}$ estimation using deep learning?
    \item What is the effect of employing different deep learning models on $T_{1\rho}$ mapping?
    \item Does zeroing out non-ROI pixels in input images improve the 2D U-Net’s $T_{1\rho}$ quantification compared to unmasked data?
\end{itemize}
\subsubsection{Experiment 1: Input Data}\label{sec:exp-input-data}

\begin{table}[b]%
\caption{$I_0$ and $I_k$ Combinations \label{tab:table2}}%
\begin{tabular*}{\columnwidth}{@{\extracolsep\fill}lll@{\extracolsep\fill}}
\toprule
\textbf{$I_0$}  & \textbf{$I_k$} & \textbf{$TSL_k$} \\
\midrule
$T_{1\rho}$-w\tnote{$^\dagger$} TSL=0ms &  $T_{1\rho}$-w & 10ms \\
$T_{1\rho}$-w TSL=0ms & $T_{1\rho}$-w & 50ms \\
PD-w\tnote{$^\ddagger$} & $T_{1\rho}$-w & 10ms \\
PD-w & $T_{1\rho}$-w & 50ms \\
\bottomrule
\end{tabular*}
\begin{tablenotes}
\item[$^\dagger$] $T_{1\rho}$-weighted image
\item[$^\ddagger$] PD-weighted FSE image
\end{tablenotes}
\end{table}

In this experiment, we investigated $T_{1\rho}$ quantification across four combinations of $I_0$ and $I_k$ in Equation \ref{eq:t1rho}. These combinations are detailed in Table \ref{tab:table2}. We selected the best-performing deep learning fitting models for each combination and compared their statistical metrics. The $T_{1\rho}$ quantification obtained using the traditional NLLS methods with these combinations served as a reference.

\subsubsection{Experiment 2: Deep Learning Model}\label{sec:exp-choice-nn}

Deep learning-based neural networks, such as U-Net and MLP, have been utilized in literature for various compositional fitting tasks. We implemented a basic U-Net and an MLP model to investigate their performance in our specific context. Both models were trained and tested on the previously mentioned four combinations of $I_0$ and $I_k$, and we aimed to identify the most effective model under each scenario.

\subsubsection{Experiment 3: ROI}\label{sec:exp-mask-roi}

\begin{figure}
\centerline{\includegraphics[width=1\linewidth]{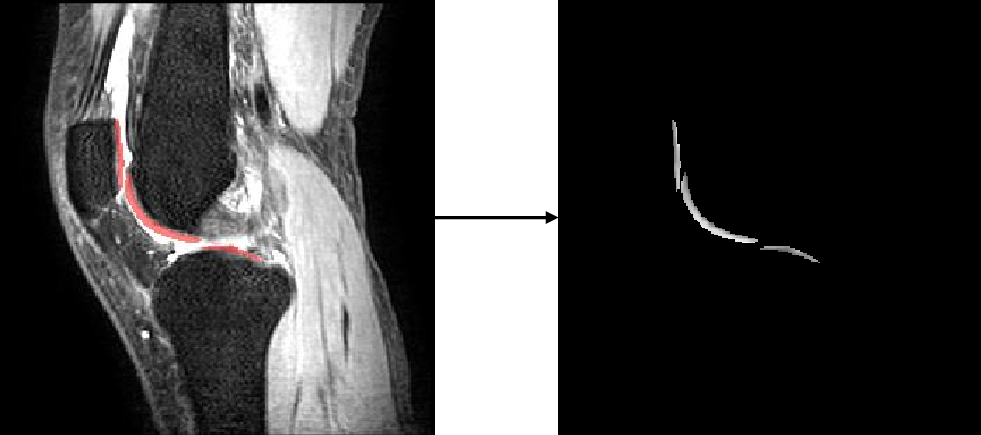}}
\caption{An illustration of masking ROI. The red overlay depicts the region extracted from this slice. This slice was obtained from the $T_{1\rho}$-weighted image (TSL=0ms) of a healthy volunteer (case V014, 22 years female, BMI = 20.76 kg/m$^2$). Note, ROI = region of interest, TSL = spin lock time, BMI = body mass index.\label{fig:masking}}
\end{figure}

Note $T_{1\rho}$-weighted images have regions that do not conform to Equation \ref{eq:t1rho}, such as areas of bone marrow. This raises the question of whether employing ROIs in the input PD-weighted and $T_{1\rho}$-weighted images to constrain the 2D U-Net could enhance the fitting performance. To investigate this, we introduced the ROI masks to zero out the voxels outside the ROI for the input images. An example of this masking operation is illustrated in Figure \ref{fig:masking}. In this context, the loss and gradient of the 2D U-Net would be computed using only the voxel intensities within the ROI, while other voxels did not participate in the optimization process. Similarly, this experiment was conducted across all four combinations of $I_0$ and $I_k$. We compared the fitting performance of the masked and unmasked 2D U-Net across all scenarios.

\subsubsection{Ground Truth and Reference}

Our ground truth was derived from a $T_{1\rho}$ quantification fitted using a conventional NLLS method with four $T_{1\rho}$-weighted images (TSL=0, 10, 30, 50ms). All metrics and experiments were evaluated against this ground truth. The authors, including a musculoskeletal radiologist (F. Xiao) with 10 years of experience, manually segmented the cartilage regions to create the cartilage ROI for various training and evaluation processes.  

We performed two-point NLLS fitting for the four combinations of $I_0$ and $I_k$ using the signal equation (Equation \ref{eq:t1rho}). This established a benchmark performance against the ground truth. By comparing this reference with our proposed methods, we were able to visualize the improvement achieved by the proposed methods under each $I_0$-$I_k$ combination.

\subsubsection{Evaluation Metrics}

We employed four evaluation metrics to statistically analyze the $T_{1\rho}$ fitting performance of the aforementioned three experiments. We trained and evaluated all deep learning models and experiments using cross-validation with the same five-fold split, while the ground truth and reference NLLS fitting were directly conducted on all subjects. The metrics were calculated at the subject level, and we reported the averages and standard deviations of each metric across the 40 subjects in our dataset.

We categorized our metrics into two types: voxel-wise and regional errors. Voxel-wise errors assess the absolute errors, while the regional errors are more aligned to the application of compositional MRI techniques such as $T_{1\rho}$, where the regional average of the quantification is typically involved.

The voxel-wise errors were assessed using two common metrics: MAE and mean absolute percentage errors (MAPE). The equations for calculating MAE and MAPE for a single sample are presented in the Equations \ref{eq:mae} and \ref{eq:mape}, where $n$ represents the number of voxels within the ROI, $\hat{y}_{i}$ denotes predicted, and $y_{i}$ corresponds to the ground truth $T_{1\rho}$ values of the voxel.

\begin{equation}\label{eq:mae}
    \text{MAE} = \frac{1}{n} \sum_{i=1}^n \left| y_i - \hat{y}_i \right|
\end{equation}

\begin{equation}\label{eq:mape}
    \text{MAPE} = \frac{1}{n} \sum_{i=1}^n \frac{\left| y_i - \hat{y}_i \right|}{\left| y_i \right|}  
\end{equation}

The regional errors were also assessed using two metrics: regional error (RE) and regional percentage error (RPE). The calculation for these metrics for one sample are presented as Equation \ref{eq:re} and \ref{eq:rpe}, where $n$ represents the number of voxels within the ROI, $\hat{y}_{i}$ denotes predicted, and $y_{i}$ corresponds the ground truth $T_{1\rho}$ values of the point.

\begin{equation}\label{eq:re}
    \text{RE} = \left| \frac{1}{n} \sum_{i=1}^n y_i - \frac{1}{n} \sum_{i=1}^n \hat{y}_i \right|
\end{equation}

\begin{equation}\label{eq:rpe}
    \text{RPE} = \frac{\left| \frac{1}{n} \sum_{i=1}^n y_i - \frac{1}{n} \sum_{i=1}^n \hat{y}_i \right|}{\left| \frac{1}{n} \sum_{i=1}^n y_i \right|}
\end{equation}

In accordance with the RSNA recommended acquisition protocol, the within-subject coefficient of variation for test-retest of cartilage $T_{1\rho}$ value ranges between 4\%-5\% \cite{chalianQIBAProfileMRIbased2021}. We translated this assertion to our target RPE, which is set to be under 5\%.

\subsection{Implementation}

The preprocessing methods were implemented using various Python \cite{vanrossumPython3Reference2009} packages, including ANTsPy \cite{avantsSymmetricDiffeomorphicImage2008} and Scikit-Image \cite{waltScikitimageImageProcessing2014}, which were utilized for image registration and preprocessing. For constructing the deep learning-based neural network, we leveraged PyTorch \cite{paszkePyTorchImperativeStyle2019}, PyTorch Lightning \cite{Falcon_PyTorch_Lightning_2019} and MONAI \cite{consortiumMONAIMedicalOpen2024}. Additionally, ITK-SNAP \cite{yushkevichUserguided3DActive2006a} was employed to prepare ROI labels, while MATLAB \cite{themathworksinc.MATLABVersion23202023}  was utilized for executing NLLS $T_{1\rho}$ fitting.

\section{Results}

\begin{figure}
\centerline{\includegraphics[width=1\linewidth]{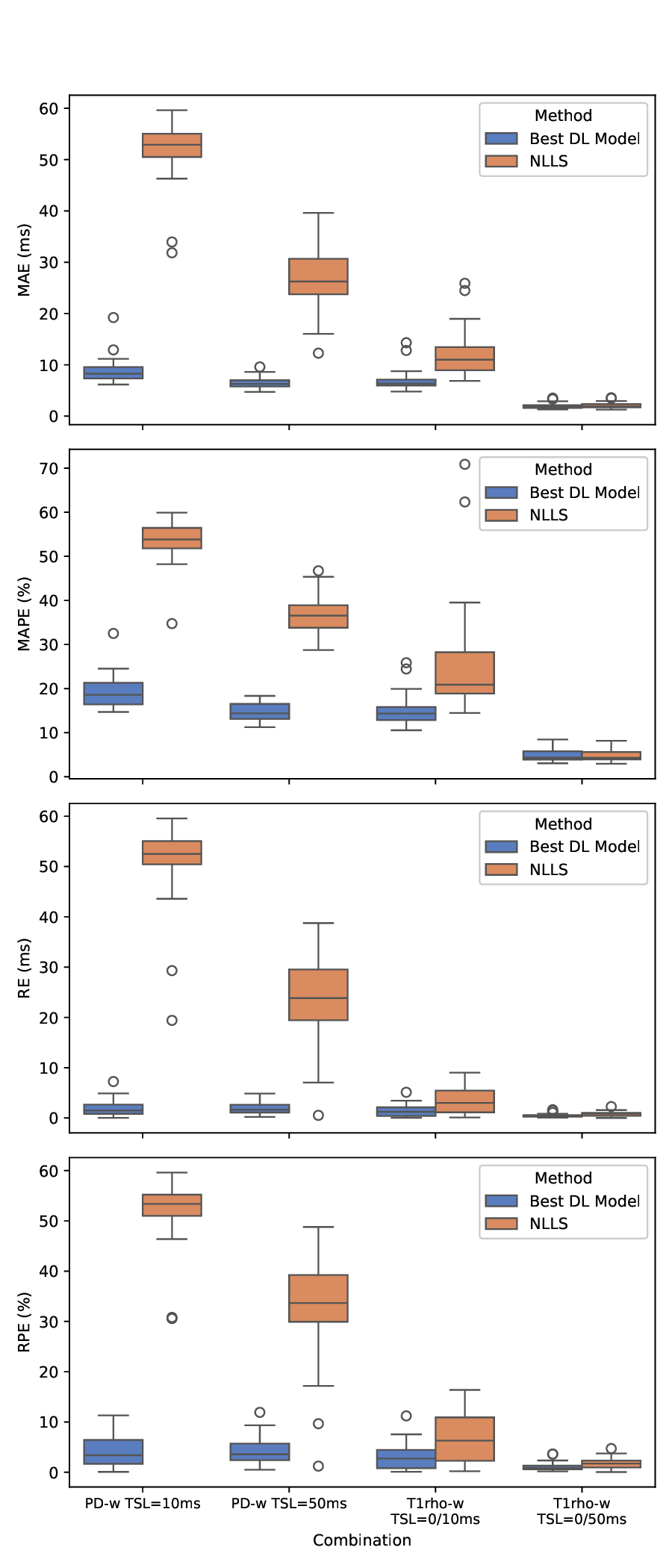}}
\caption{Boxplots comparing the best deep learning models and NLLS fitting on all $I_0$-$I_k$ combinations. Note, DL = deep learning, NLLS = non-linear least square, MAE = mean absolute error, MAPE = mean absolute percentage error, RE = regional error, RPE = regional percentage error, PD-w = proton density-weighted image, TSL = time of spin-lock, T1rho-w = $T_{1\rho}$-weighted image.\label{fig:dl-nlls}}
\end{figure}

\begin{figure*}
\centerline{\includegraphics[width=1\linewidth]{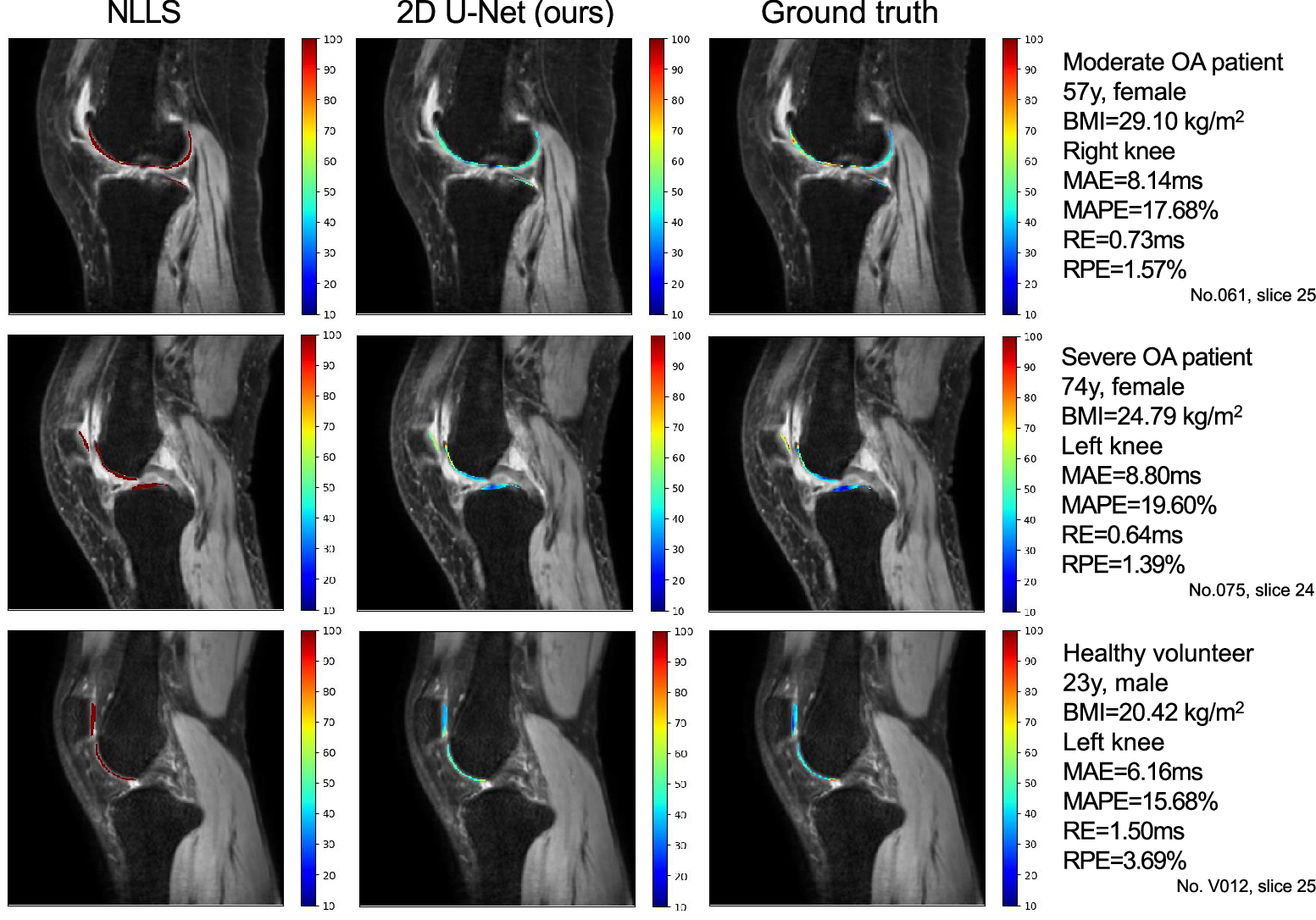}}
\caption{Example results (middle column) of fitting $T_{\rho}$ from PD-weighted and $T_{1\rho}$-weighted image (TSL=10ms) using a 2D U-Net (not masking ROI). The left and right columns present the reference and ground truth results fitted by NLLS. The reference NLLS $T_{1\rho}$ was fitted by the aforementioned two images, whereas the ground truth was derived from four $T_{1\rho}$-weighted images. Note, NLLS=non-linear least square fitting, OA=osteoarthriths, BMI=body mass index, MAE=mean absolute error, MAPE=mean absolute percentage error, RE=regional error, RPE=regional percentage error, PD=proton density, TSL=time of spin-lock, ROI=region of interest.}\label{fig:sample_results}
\end{figure*}

\begin{table*}[t]%
\caption{Experiment results\label{tab:results}}
\begin{tabular*}{\textwidth}{@{\extracolsep\fill}llcccc@{\extracolsep\fill}}
\toprule
Model & Metrics\tnote{$^{1}$} & \makecell{PD-w\tnote{$^{2}$} \\ TSL\tnote{$^{3}$}=10ms} & \makecell{PD-w \\ TSL=50ms} & \makecell{TSL=0ms \\ TSL=10ms} & \makecell{TSL=0ms \\ TSL=50ms} \\
\midrule
% 2D U-Net / unmasked
\multirow{4}{*}{\makecell{2D U-Net \\ unmasked}} & MAE (ms) & \textbf{8.72 $\pm$ 2.26} & 6.53 $\pm$ 1.04 & \textbf{6.82 $\pm$ 1.83} & 3.87 $\pm$ 0.52 \\
& MAPE (\%) & 19.00 $\pm$ 3.39 & 14.73 $\pm$ 2.10 & \textbf{14.94 $\pm$ 3.19} & 9.19 $\pm$ 1.42 \\
& RE (ms) & \textbf{1.93 $\pm$ 1.54} & \textbf{1.84 $\pm$ 1.15} & \textbf{1.42 $\pm$ 1.15} & 0.83 $\pm$ 0.60 \\
& RPE (\%) & \textbf{4.12 $\pm$ 3.01} & \textbf{4.03 $\pm$ 2.63} & \textbf{3.15 $\pm$ 2.61} & 1.79 $\pm$ 1.30 \\
\hline
% 2D U-Net / masked
\multirow{4}{*}{\makecell{2D U-Net \\ masked}} & MAE (ms) & 9.06 $\pm$ 3.35 & \textbf{6.49 $\pm$ 1.83} & 7.01 $\pm$ 2.81 & 3.15 $\pm$ 1.65 \\
& MAPE (\%) & 22.62 $\pm$ 24.23 & 17.22 $\pm$ 18.92 & 17.51 $\pm$ 18.66 & 10.85 $\pm$ 19.53 \\
& RE (ms) & 3.91 $\pm$ 3.72 & 3.03 $\pm$ 2.02 & 2.58 $\pm$ 2.84 & 0.85 $\pm$ 1.68 \\
& RPE (\%) & 8.28 $\pm$ 6.90 & 6.70 $\pm$ 4.70 & 5.45 $\pm$ 5.68 & 2.12 $\pm$ 4.96 \\
\hline
% 1D MLP
\multirow{4}{*}{\makecell{1D MLP\tnote{$^{4}$}}} & MAE (ms) & 9.54 $\pm$ 3.63 & 6.89 $\pm$ 2.29 & 7.73 $\pm$ 3.08 & \textbf{1.97 $\pm$ 0.51} \\
& MAPE (\%) & \textbf{18.91 $\pm$ 3.90} & \textbf{14.05 $\pm$ 3.38} & 15.07 $\pm$ 3.96 & \textbf{4.84 $\pm$ 1.27} \\
& RE (ms) & 4.60 $\pm$ 4.53 & 3.80 $\pm$ 3.22 & 3.41 $\pm$ 3.17 & \textbf{0.48 $\pm$ 0.33} \\
& RPE (\%) & 9.39 $\pm$ 7.56 & 8.08 $\pm$ 6.61 & 6.96 $\pm$ 5.18 & \textbf{1.05 $\pm$ 0.77} \\
\hline
% NLLS
\multirow{4}{*}{\makecell{NLLS\tnote{$^{5}$}}} & MAE (ms) & 52.20 $\pm$ 5.64 & 26.83 $\pm$ 6.16 & 11.99 $\pm$ 4.28 & 2.04 $\pm$ 0.54 \\
& MAPE (\%) & 53.54 $\pm$ 4.40 & 36.54 $\pm$ 4.11 & 24.89 $\pm$ 11.49 & 4.74 $\pm$ 1.19 \\
& RE (ms) & 51.41 $\pm$ 7.46 & 23.69 $\pm$ 8.43 & 3.39 $\pm$ 2.42 & 0.81 $\pm$ 0.48 \\
& RPE (\%) & 52.35 $\pm$ 6.10 & 32.99 $\pm$ 9.71 & 6.76 $\pm$ 4.55 & 1.74 $\pm$ 1.05 \\
\bottomrule
\end{tabular*}
\begin{tablenotes}%%[341pt]
\item The results were shown as mean $\pm$ standard deviation among the samples. The \textbf{bold} text shows the best model in a given metrics and data combo. The header row of the last four columns shows the data combos, $I_0$ at the top and $I_k$ at the bottom line. TSL=\textit{x} ms represents a $T_{1\rho}$-weighted image prepared by the given \textit{x} ms of TSL.
\item[$^{1}$] MAE = mean absolute error, MAPE = mean absolute percentage error, RE = regional error, RPE = regional percentage error.
\item[$^{2}$] PD-w = proton density (PD)-weighted FSE MRI.
\item[$^{3}$] TSL = spin lock time.
\item[$^{4}$] MLP = multi-layer perceptron. 
\item[$^{5}$] NLLS = non-linear least square.
\end{tablenotes}
\end{table*}

We presented our comprehensive result in Table \ref{tab:results}. For the sake of clarity and organization, we have categorized our interpretations according to the experimental framework. Our analysis indicates that across all $I_0$-$I_k$ combinations, the RPE of the best-performing deep learning models consistently remained below our predefined threshold of 5\%.

\subsection{Experiment 1: Input Data}

We conducted a comparative analysis of the top-performing models within each $I_0$-$I_k$ combination alongside the reference NLLS. The boxplots for all four evaluation metrics are illustrated in Figure \ref{fig:dl-nlls}. Notably, deep learning models consistently outperformed the NLLS reference with lower mean and standard deviations of errors. However, the performance disparity between the two methods increased in challenging scenarios, such as PD-weighted and $T_{1\rho}$-weighted collected with TSL=10ms. Conversely, the reference NLLS fitting exhibited significantly reduced errors, indicating a comparable performance to the best deep learning model when processing $T_{1\rho}$-weighted images with TSLs of 0ms and 50ms.

We identified the MLP as the top-performing deep learning fitting model utilizing two $T_{1\rho}$-weighted images when TSLs were 0ms and 50ms, achieving an RPE of 1.05$\pm$0.77\%, mean$\pm$standard deviation. This was followed by the unmasked 2D U-Net model trained on two $T_{1\rho}$-weighted images when TSLs were 0ms and 10ms, which yielded an RPE of 3.15$\pm$2.61\%. In scenarios where $I_0$ was a PD-weighted image, the best models remained the unmasked 2D U-Net, which exhibited a performance decline compared to the $T_{1\rho}$ cases mentioned earlier, resulting in RPEs of 4.03$\pm$2.63\% and 4.12$\pm$3.01\% when $I_k$ corresponded to $T_{1\rho}$-weighted images with TSLs were 50ms and 10ms, respectively. Although the RPEs were comparable, we observed a lower MAE (6.49$\pm$1.83ms) when processing $T_{1\rho}$-weighted images with TSL=50ms. To provide further insight, we selected example slices to display the fitting results from the best model, the NLLS reference, and the ground truth in Figure \ref{fig:sample_results}.

\subsection{Experiment 2: Deep Learning Model}

A comparative analysis of the unmasked 2D U-Net and the MLP models was conducted in this experiment, focusing on four distinct combinations of $I_0$ and $I_k$. Notably, it was observed that the MLP model (RPE=1.05$\pm$0.77\%) outperformed the unmasked 2D U-Net (RPE=1.79$\pm$1.30\%) exclusively when $I_0$ and $I_k$ corresponded to $T_{1\rho}$-weighted images with TSLs of 0ms and 50ms. Conversely, in all other scenarios, the unmasked 2D U-Nets exhibited superior performance compared to the MLP model. These results are presented in the second and fourth rows in Table \ref{tab:results}.

\subsection{Experiment 3: ROI}

In this experiment, a comparative study was undertaken to investigate the effects of masking the ROI on the performance of both masked and unmasked 2D U-Nets in four distinct combinations of $I_0$ and $I_k$. The results revealed that masking the ROI consistently degraded the fitting performance in all scenarios, particularly in terms of RPE. Notably, none of the models were able to surpass the performance of the unmasked U-Net and MLP within the same $I_0$-$I_k$ combinations. These findings are presented in the first and second rows of Table \ref{tab:results}.

\section{Discussion}

Our study used deep learning techniques to investigate the potential of utilizing PD-weighted anatomical FSE images for $T_{1\rho}$ quantification. To evaluate this approach, we conducted various experiments on a dataset comprising OA patients and healthy volunteers, examining the performance of deep learning fitting methods under different input data, deep learning model architectures, and preprocessing settings. Our results demonstrated that we achieved the best RPE of less than 5\% across all four $I_0$ and $I_k$ combinations, as detailed in Table \ref{tab:table2}, thereby substantiating our hypothesis and establishing a reliable fitting method \cite{chalianQIBAProfileMRIbased2021}.

In this study, we investigated the selection of $TSL_k$ (in the $T_{1\rho}$ signal Equation \ref{eq:t1rho}) using the corresponding $I_k$ images. In spin-lock-based acquisitions, the maximum TSL is limited by the RF amplifier configurations and SAR \cite{roeloffsQuantificationPulsedSpinlock2015}. We noted that a previous study suggested a shorter $TSK_k$ lead to poorer $T_{1\rho}$ fitting performance \cite{huangUncertaintyawareSelfsupervisedNeural2022}, and our result demonstrated a similar trend when $TSL_k$ was reduced. Nevertheless, the obtained $T_{1\rho}$ still satisfied the recommended standard (RPE less than 5\%)\cite{chalianQIBAProfileMRIbased2021}. This finding further supports the feasibility of reducing TSL of the $T_{1\rho}$ protocols, alleviating scanner hardware and SAR requirements

Our experiments were conducted to elucidate the efficacy of deep learning methods under various data input scenarios and model configurations. We aimed to identify which model performed optimally with a given data input. In our proposed scenario of $T_{1\rho}$ fitting, the deep learning models, namely neural networks, operated as universal approximators \cite{hornikMultilayerFeedforwardNetworks1989} of the $T_{1\rho}$ signal equation (Equation \ref{eq:t1rho}). It was anticipated that these neural networks would effectively fit $T_{1\rho}$ from $I_0$ and $I_k$, exhibiting comparable performance to the standard NLLS fitting under normal conditions, i.e., multiple $T_{1\rho}$-weighted images. Nonetheless, mathematically, we were tasked with fitting $T_{1\rho}$ from the minimum number of images, where the algorithms must balance the noise and out-of-distribution (OOD) signals to achieve accurate fitting.

Our comparisons with two classic deep learning models, the 2D U-Net and the 1D MLP, revealed that the 2D U-Net was more adept at addressing the challenges posed by noise and OOD data. This assertion is substantiated by the trend analysis of metrics against the data input combinations ($I_0$-$I_k$): the 2D U-Net outperformed the 1D MLP and NLLS fitting in all combinations involving PD-weighted images and $T_{1\rho}$-weighted images collected at a TSL of 10 ms. Conversely, the 1D MLP surpassed the performance of the 2D U-Net and reference NLLS fitting in the remaining combinations. We propose that this distinction arose due to the 2D U-Net's capacity to capture spatial information from the input patches \cite{krizhevskyImageNetClassificationDeep2017}, which a standard 1D MLP lacked. Moreover, the convolution operation within the 2D U-Net facilitated the removal of noise signals \cite{zhang2023dive}. These attributes endowed the 2D U-Net with a robust ability to leverage additional information and balance the OOD data for accurate $T_{1\rho}$ quantification. Conversely, the 1D MLP approximated the $T_{1\rho}$ signal equation more effectively in scenarios characterized by minimal noise and OOD signals.

Following the universal approximation theorem \cite{hornikMultilayerFeedforwardNetworks1989}, and the assumption related to $T_{1\rho}$ imaging, we experimented with masking the region outside the cartilage ROI for the 2D U-Net. We anticipated that the 2D U-Net would concentrate more on the region that follows the $T_{1\rho}$ signal equation (Equation \ref{eq:t1rho}). However, our results indicated a negative influence from this masking approach. As CNN-based models like U-Net \cite{ronnebergerUNetConvolutionalNetworks2015} are designed to capture and encode spatial information from the training data \cite{krizhevskyImageNetClassificationDeep2017, zhang2023dive}, we interpreted this performance decrease as indicative of the importance of voxels outside ROI for the 2D U-Net to achieve accurate $T_{1\rho}$ quantification for the ROI. Although the intensities of those voxels neither adhere to the signal equation (Equation \ref{eq:t1rho}) nor contribute directly to the fitting task itself, they still provide valuable anatomical information about the knee and spatial features when considered in the context of image processing.

$T_{1\rho}$ imaging holds great potential as a diagnostic tool for cartilage assessment; however, its limited adoption in clinical practice highlights the need for strategies to integrate it into routine workflows. One promising approach is to derive $T_{1\rho}$ maps directly from conventional anatomical images, such as clinical fast spin echo (FSE) sequences, which are commonly used for knee imaging. This method could address the challenge of additional scan time required for $T_{1\rho}$ MRI by incorporating $T_{1\rho}$ quantification into standard clinical protocols. Previous studies support this strategy: Santyr et al. demonstrated that Carr-Purcell-Meiboom-Gill (CPMG) acquisitions can replicate spin-locking behavior using density matrix theory when the echo spacing matches the spin-lock frequency \cite{santyrVariationMeasuredTransverse1988}, and Gold et al. demonstrated this in in vivo experiments \cite{goldT1rImagingArticular2006}. The ultimate goal is to derive $T_{1\rho}$ values entirely from standard clinical protocols without requiring additional acquisitions at non-zero TSL, enabling simultaneous anatomical and biochemical assessment of knee cartilage within routine imaging workflows.

This study presents several limitations. First, it is a retrospective investigation, and the acquisition parameters of the PD-weighted FSE and $T_{1\rho}$-weighted scans were not optimized for the proposed approach. Although registration and meticulous ROI labeling were employed to mitigate this limitation, these measures only partially addressed the variations, as evidenced by the substantial standard deviation observed in the results. Furthermore, the absence of validation through longitudinal studies restricts the ability to confirm the OA detection performance of our method. To overcome these limitations, future work will concentrate on designing a tailored acquisition protocol and validating the approach through comprehensive longitudinal studies.

In summary, we have proposed accelerated acquisition and processing methods for knee $T_{1\rho}$ quantification using PD-weighted and $T_{1\rho}$-weighted FSE MRI alongside deep learning. We demonstrated our approach on a dataset collected from OA patients and healthy volunteers. Notably, our approach exhibits compatibility with $T_{1\rho}$-weighted images acquired with a shorter TSL of 10ms, thereby enabling the reduction of RF power and SAR during optimized $T_{1\rho}$ acquisition. To further refine our method, we conducted experiments to explore the effects of various design choices, providing valuable insights into selecting optimal deep learning models for fitting $T_{1\rho}$ in clinical MRI. The proposed approach has the potential to facilitate the incorporation of advanced quantitative MRI methods into routine clinical practice, ultimately benefiting patients and the broader population.

\section*{Acknowledgments}

This work was supported by a grant from the Innovation and Technology Commission of the Hong Kong SAR [MRP/001/18X]. We would like to acknowledge Chi Yin Ben Choi and Cheuk Nam Cherry Cheng for their assistance in patient recruitment and MRI exams and Tsz Shing Adam Kwong for his assistance in data processing. We used Large Language Models, including a self-hosted Llama 3 (Meta AI) and an online GPT-4o Mini (Open AI) service hosted by Poe (Quora Inc.) for editing the text of the manuscript in December 2024 and January 2025.

\subsection*{Financial disclosure}

None reported.

\subsection*{Conflict of interest}

The authors declare no potential conflict of interest.

\bibliography{MRM-AMA}%
\vfill\pagebreak

% \nocite{*}% Show all bib entries - both cited and uncited; comment this line to view only cited bib entries;

\end{document}